\documentclass[12pt]{article}
\usepackage{epsfig}
\usepackage{graphics,amssymb}
\begin{document}

\thispagestyle{empty}
\renewcommand{\thefootnote}{\fnsymbol{footnote}}

\begin{flushright}
{\normalsize
SLAC-PUB-8846\\
May 2001}
\end{flushright}

\vspace{.8cm}

\begin{center}
{\bf\Large
Impedance Analysis of Bunch Length Measurements at the ATF Damping
Ring}\footnote{Work supported by Department of Energy
contract DE--AC03--76SF00515, and by the Chinese National Foundation
of Natural Sciences, contract 19975056-A050501.}

\vspace{1cm}

{\large
K.L.F. Bane,
T. Naito\footnote[3]{High Energy Accelerator Research Organization (KEK),
1-1 Oho, Tsukuba, Ibaraki, Japan.}
, T. Okugi\footnotemark[3], Q. Qin\footnote[4]{Institute of High
Energy Physics (IHEP), Beijing, People's Republic of China.}, and J. Urakawa\footnotemark[3]\\
{\it Stanford Linear Accelerator Center, Stanford University,}\\
{\it Stanford, CA 94309 USA.}}

\end{center}

\vspace{.5cm}

\abstract{
We present energy spread and bunch length measurements at the Accelerator Test
Facility (ATF) at KEK, as functions of current, for different ring rf
voltages, and with the beam both on and off the coupling resonance. We fit the
on-coupling bunch shapes to those of an impedance model consisting of a resistor
and an inductor connected in series. We find that the fits are reasonably good, but that
the resulting impedance is unexpectedly large.
}

\vfill

\begin{center}
{\it Presented at the}
{\it 10$^{\it th}$ International Symposium on Applied Electromagnetics and Mechanics (ISEM2001)} \\
{\it Toshi Center Hotel, Tokyo, Japan}\\
{\it May 13-16, 2001}
\end{center}

\newpage

\section{Introduction}

In future e+e- linear colliders, such as the JLC/NLC, damping
rings are needed to generate beams of intense bunches with very
low emittances. A prototype for such damping rings is the
Accelerator Test Facility (ATF)\cite{ATF} at KEK. One important
consideration for such rings is that  the (broad-band)
longitudinal impedance be kept sufficiently small, to avoid
(longitudinal) emittance growth caused by potential well
distortion and/or the microwave instability. Measurements of
energy spread and bunch length as functions of current are a way
of verifying the size of the impedance and its effects.
The
ATF, as it is now---running below design energy and with the wigglers
turned off---is strongly affected by
intra-beam scattering (IBS), an effect that modifies all
dimensions of the beam.
To
study the impedance effects alone, however, the machine can be run
on a (difference) coupling resonance, where the vertical beam size
grows and the IBS forces become weak.

Calculations of the impedance of the ATF
ring vacuum chamber
yield a total inductive component (at the typical
bunch lengths) of $\sim15$~nH\cite{Kim};
the dominant resistive component, the rf cavities,
are expected to contribute $\sim100$~$\Omega$.
To obtain the impedance from bunch shape
measurements, if the data is noisy (as it will be in our case), we
need to fit to a relatively simple model of impedance.
A pure resistor\cite{Holtz}, a pure inductor\cite{oldSLC}, and
a broad-band resonator\cite{resonator} are all impedance models that have been
used to characterize the impedance of storage rings. A simple model
that can account for both potential well bunch lengthening and parasitic mode losses
is a resistor and an inductor connected in series. This model was
used to analyze bunch length measurements at CESR, where it appeared to
fit the data well\cite{CESR}. In this report we present bunch length measurements
at the ATF, we use this model to estimate the real and imaginary parts of
the ATF impedance, and then we compare our results with the earlier estimates.

\section{Measurements}

\begin{figure}[htb]
\centering
\epsfig{file=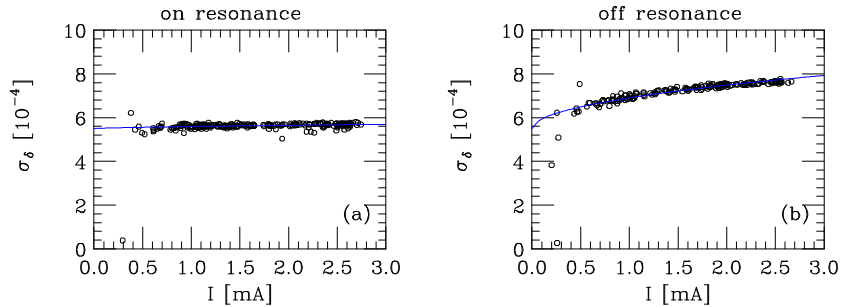, height=4cm}
\caption{
Energy spread as function of current when the ring voltage $V_c=300$~kV, with the beam
on~(a) and off~(b) the coupling resonance.
}
\label{fisige}
\end{figure}

To obtain the beam energy spread in the ATF damping ring, the beam width is measured
after extraction on a thin screen in a region of high dispersion. In Fig.~\ref{fisige} we
plot the measured rms energy spread $\sigma_\delta$ {\it vs.} current $I$ for peak rf
voltage $V_c=300$~kV,
for the case of the beam on resonance~(a), and off~(b).
We note that, by $I=2.5$~mA, on the coupling
resonance the energy spread growth is still very small (3\%),
whereas off resonance it is not (36\%). Note also that Fig.~\ref{fisige}a
implies that the threshold to the microwave instability---whose signature would be
a kink in the data---must be beyond $I=2.7$~mA.

The bunch length in the ATF ring was measured using a Hamamatsu C5680 streak camera.
The data taking process consists of storing a high current beam, and then measuring the
longitudinal bunch profile 50-70 times
at fixed time intervals, while the current naturally decreased.
The profiles were stored to disk, along with their DCCT current monitor readings.
The measurements were
repeated for $V_c=150$, 200, 250, 300~kV, and for the beam both on and off the
coupling resonance. Each trace was fit to an asymmetric Gaussian composed of two
half-Gaussians with lengths $\sigma(1\pm\epsilon)$, where $\sigma$ is the rms bunch
length and $\epsilon$ the asymmetry
parameter.
 The parameters
$\sigma$ ($\epsilon$)
gives us information about the imaginary (real) part of the impedance.
Note that for the asymmetric Gaussian $\epsilon\approx.63z_{skew}$, where
$z_{skew}$ is the skew moment of the distribution.
Note also that we expect the bunch to lean forward, which, in our convention,
will mean $\epsilon>0$.
Instead of the 3rd moment we would prefer to use the 1st moment of the
bunch distribution to probe the real part of the impedance;
the streak camera trigger,
however, is not accurate enough to resolve the kind of centroid shifts
needed (on the order of picoseconds).

\begin{figure}[htb]
\centering
\epsfig{file=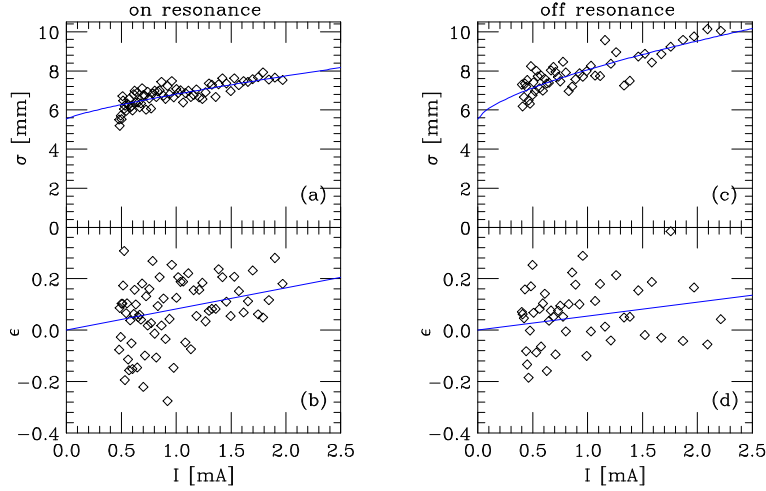, height=6.5cm}
\caption{
Bunch length ($\sigma$) and asymmetry parameter ($\epsilon$) of the asymmetric
Gaussian fit to the measured bunch distributions, as functions of current,
for the beam on resonance (a,b) and off (c,d).
$V_c=250$~kV. The curves are fits to these results.
}
\label{fisigz}
\end{figure}

Results for $V_c=250$~kV are shown in Fig.~\ref{fisigz}.
We note that $\sigma$ increases with current, even on resonance
(at 2mA the growth is 40\%), implying that there is significant potential well
distortion in the ATF. Off-resonance, however, the growth is much larger (at 2mA, 72\%),
due to IBS. As for the asymmetry parameter,
we see much scatter in the data.
We fit the $\epsilon$ results with a straight line through the origin
(see Fig.~\ref{fisigz}b,d).
On-resonance we find that $\epsilon\approx .1I/$mA, and it decreases slightly with $V_c$.
Note that for all eight sets of measurements
(four voltages, both on and off the coupling resonance)
the slopes are positive, as we expect from physical considerations.
Thus, despite the large scatter in the data, there appears to be
physical information in the fitted $\epsilon$ (or equivalently, the skew moment of the
distribution) that we can draw out through statistics.
Finally,
note that more details of the bunch length measurements can be found in Ref.\cite{ATFsigz}.

\section{Ha\"{\i}ssinski Solution for an $R+L$ Impedance}

To obtain the steady-state bunch distribution for a series $R+L$ impedance
we begin with
the Ha\"{\i}ssinski equation\cite{Haiss}:
\begin{equation}
\lambda(z)={\exp\left( -{z^2\over 2\sigma_{z0}^2} + {1\over
  V^\prime_{rf}\sigma_{z0}^2}\int_0^z V_{ind}(z^\prime)\,dz^\prime\right)\over
  \int_{-\infty}^\infty \exp\left( -{z^2\over 2\sigma_{z0}^2} + {1\over
  V^\prime_{rf}\sigma_{z0}^2}\int_0^z V_{ind}(z^\prime)\,dz^\prime\right) dz}
  \quad,
  \label{eqHai}
\end{equation}
where the induced voltage is given by
\begin{equation}
V_{ind}(z)= -\int_0^\infty
W(z^\prime)\lambda(z-z^\prime)\,dz^\prime\quad,
\end{equation}
with
$\lambda$ the bunch position distribution, $z$ longitudinal position
($z<0$ is toward the front of the bunch),
$V_{rf}^\prime$ the slope of the rf voltage at the synchronous point,
$\sigma_{z0}$ the nominal (zero current) bunch length, and
$W(z)$ the (point charge) wakefield. Note that
$\int_{-\infty}^\infty\lambda(z)dz=1$.
 Eq.~\ref{eqHai} can be written
as a first order, non-linear differential equation
\begin{equation}
{\lambda^\prime\over\lambda}= -{z\over\sigma_{z0}^2} +
{V_{ind}(z)\over V_{rf}^\prime\sigma_{z0}^2}\quad. \label{eqdiff}
\end{equation}
For the special case of a resistive plus inductive impedance in series,
with resistance $R$ and inductance $L$,
\begin{equation}
V_{ind}=-eNc(R\lambda+cL\lambda^\prime)\quad.
\end{equation}
For this case, Eq.~\ref{eqdiff} can be written in normalized units
as
\begin{equation}
y^\prime=-y{ (x+ry)\over(1+\ell y) }\quad, \label{eqdiffn}
\end{equation}
where $x=z/\sigma_{z0}$, $y(x)=\lambda\sigma_{z0}$.
The normalized induced voltage $v=V_{ind}/(V_{rf}^\prime\sigma_{z0})$.
Note that
there are two free parameters in our equation: the (normalized)
resistance times current,
 $r=ecNR/(V_{rf}^\prime\sigma_{z0}^2)$,
 and the (normalized) inductance times
current,
 $\ell=ec^2NL/(V_{rf}^\prime\sigma_{z0}^3)$.
To solve Eq.~\ref{eqdiffn}, we begin at a position $x_0$ far in front
of the bunch, choose $y(x_0)$ (a small number), numerically solve the
differential equation, and compute the total integral of $y(x)$.
We then iterate this process, adjusting $y(x_0)$ until the
integral equals 1.

\begin{figure}[htb]
\centering
\epsfig{file=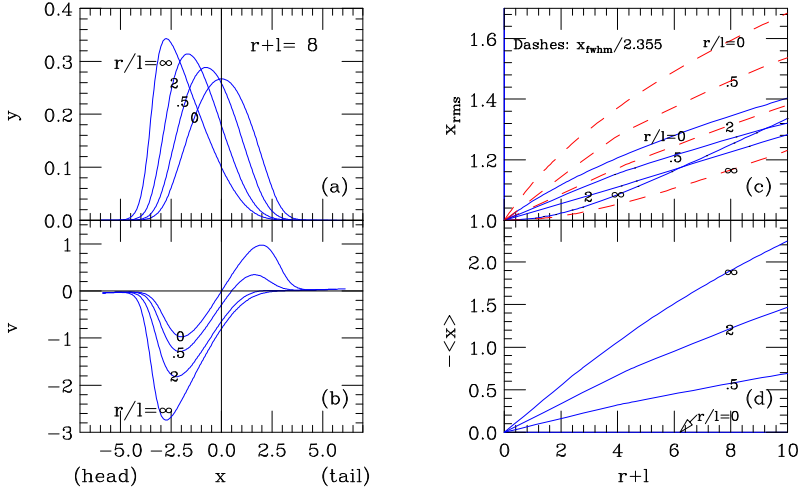, height=6.5cm}
\caption{
For a series $R+L$ impedance, the bunch shape~(a) and the induced voltage~(b)
for example cases when $r+\ell=8$; the rms length of the distribution~(c)
and the centroid shift~(d)as functions of $(r+\ell)$. The dashes
in (c) give $x_{fwhm}/2.355$.
}
\label{fihaiss}
\end{figure}

Numerical results are shown in Fig.~\ref{fihaiss}.
In (a,b) we show representative bunch shapes and induced voltages for the
example $r+\ell=8$;
in (c,d) we give the first and second moments of the bunch shape,
and also the full-width-at-half-maximum $z_{fwhm}$,
as functions of $(r+\ell)$.

\section{Fitting to Bunch Length Measurements}

\begin{figure}[htb]
\centering
\epsfig{file=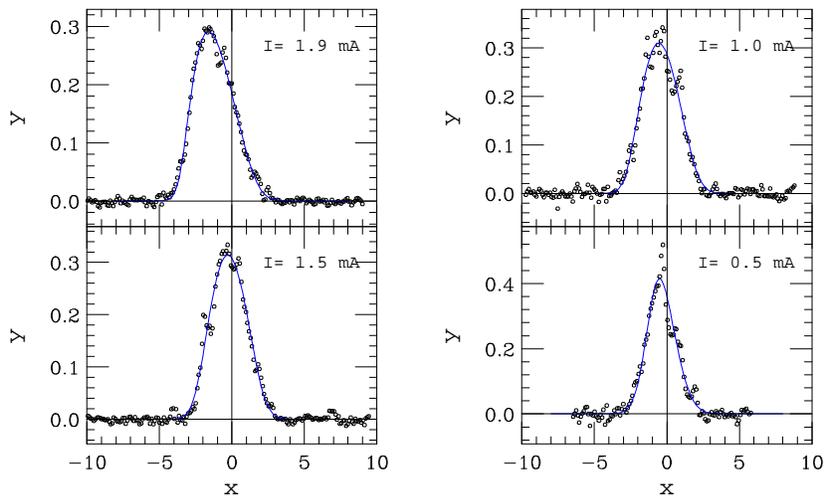, height=6.5cm}
\caption{
Four measured bunch shapes and their fits to the
Ha\"{\i}ssinski solution of a series $R+L$ impedance. Here $V_c=250$~kV.
Note that the head of the beam is to the left.
}
\label{fi1350ex}
\end{figure}

We fit the on-coupling bunch length measurements to
$a_0+a_1x+a_2y_H(r,\ell,x_{shift},\sigma_{z0})$, where
$y_H$ is the Ha\"{\i}ssinski solution to the $R+L$ impedance.
The fitting parameters are $a_0$, $a_1$, $a_2$, $r$, $\ell$, and
$x_{shift}$, a centroid shifting parameter.
We use the Method of Maximum
Likelihood to do the error analysis, a method that assumes
all errors are purely random and normally distributed\cite{Bev}.
Fig.~\ref{fi1350ex} shows four sample fits (here
$x$ is in the unshifted Ha\"{\i}ssinski frame). We see some scatter
in the data, though the fits are reasonably good.

\begin{figure}[htb]
\centering
\epsfig{file=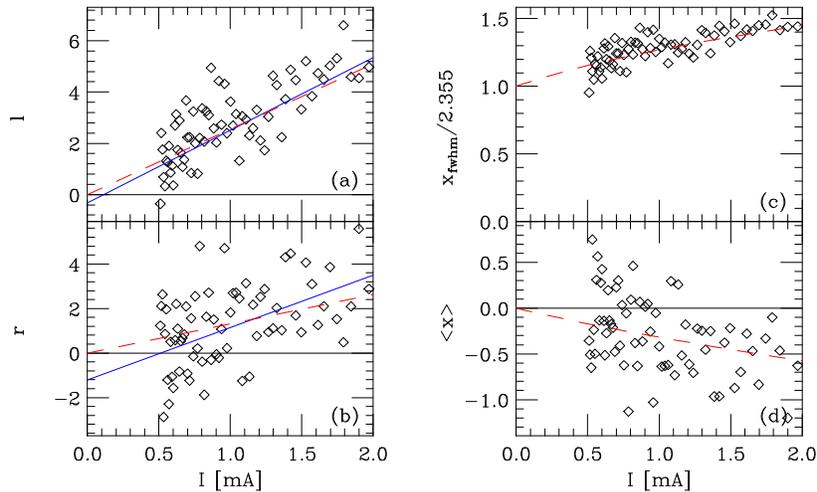, height=6.5cm}
\caption{
Fitting results for $V_c=250$~kV: the fitting parameters $\ell$~(a), $r$~(b); for the
fitted distributions, the full-width~(c) and the centroid position~(d).
}
\label{fi1350summ}
\end{figure}

In Fig.~\ref{fi1350summ} we summarize the fits for
 $V_c=250$~kV. Shown are the fitting parameters $\ell$~(a) and $r$~(b);
the full-width~(c) and centroid position~(d) of the fitted distributions.
In (a) and (b) the solid lines are least squares fits to the results,
the dashed lines are least squares fits through the origin.
We see that the least squares fit for $\ell$ naturally passes very close to
the origin; the fit for $r$, however, does not pass so close, indicating that
the resistive part of the model does not fit as well.
(This observation is also true for the cases of the other rf voltages.)
Nevertheless, as
our solution we take the least squares fit through the origin
for both parameters (the dashes).
The resulting full-width and centroid shift are given by the dashes
in (c,d).
We should note also that the fitted values of $a_2$, a parameter that
should be proportional to the bunch charge, correlated very well with the
actual DCCT current readings.

In Table 1 we give a summary of the fitting results for the four different
voltages.
From the scatter in the entries
the overall results are estimated to be: $R=1.25\pm.35$~k$\Omega$
and $L=44.5\pm7.5$~nH.
When the measurements were repeated six months later,
and a slightly different analysis was applied, the results were:
$R=1.65\pm.20$~k$\Omega$ and $L=32.5\pm1.0$~nH.

\begin{table}[htb]
\centering
\caption{Fitting Summary.}
\vspace*{2mm}
\begin{tabular}{||c|c|c||}\hline\hline
$V_c$ (kV) & $R$ (k$\Omega$) & $L$ (nH) \\ \hline\hline
150 & 1.3$\pm$.05 & 54$\pm$.7 \\ \hline
200 & 1.7$\pm$.05 & 50$\pm$.6 \\ \hline
250 & 1.1$\pm$.04 & 39$\pm$.4 \\ \hline
300 & 0.8$\pm$.04 & 35$\pm$.4 \\ \hline\hline
\end{tabular}
\end{table}

\section{Discussion}

From Table 1 we note that the scatter in the results, for both $R$ and $L$,
is much larger than the estimated errors, implying that there are significant
systematic errors or problems with
the model. Systematic errors might include problems with the streak
camera, such as residual space charge or aberrations in the optics.
Error in the rf voltage, and consequently in the parameter
$\sigma_{z0}$, can also contribute to a systematic error in the
results, though we don't think that this a significant effect.
To improve the fit to the data (remember: the least squares fit to
the fitted $r(I)$ deviated from the origin)
a different impedance model, such as a broad-band resonator, can also be tried.
Nevertheless, the fits to our model were fairly good and
the results are reasonably consistent, at least to $\pm$(20-30)\%.

The impedance values implied by our analysis
are much larger than expected: $L$ is about a factor
of 3 larger than earlier calculations, and $R$ is about a factor of 10 larger,
a result that is especially puzzling.
To see whether there actually is such a large, resistive impedance
in the ATF, it is planned, in the near future, to directly measure the beam
synchronous phase shift with current using a synchroscan streak camera,
one whose timing is accurately tied to the rf system timing.
With such hardware, we believe that the expected 2-3 degree
phase shift (at 714~MHz) in going from high to low current
should be relatively easy to detect.


\end{document}